\documentclass[a4paper]{iopconfser}

\usepackage{amssymb}
\usepackage{amsmath}
\usepackage{amscd}
\usepackage{latexsym}
\usepackage{graphicx}
\usepackage{cite}

\newcommand{\be}{\begin{equation}}
\newcommand{\ee}{\end{equation}}
\newcommand{\Dlt}{\Delta}
\newcommand{\dlt}{\delta}
\newcommand{\prt}{\partial}
\newcommand{\bfr}{{\bf r}}

\newcommand{\bt}{\beta}
\newcommand{\vp}{\varphi}
\newcommand{\ep}{\varepsilon}
\newcommand{\al}{\alpha}

\newcommand{\om}{\omega}
\newcommand{\Om}{\Omega}

\newcommand{\dgr}{\dagger}

\newcommand{\rgl}{\rangle}
\newcommand{\lgl}{\langle}
\newcommand{\cH}{{\cal H}}
\newcommand{\cL}{{\cal L}}

\begin{document}

\title{Nonlinear coherent modes and atom optics} 

\author{V.I. Yukalov$^{1,2}$, E.P. Yukalova$^{3}$, and V.S. Bagnato$^2$ } 

\affil{
$^1$Bogolubov Laboratory of Theoretical Physics, 
Joint Institute for Nuclear Research, \\
Dubna 141980, Russia} 

\affil{
$^2$Instituto de Fisica de S\~ao Carlos, Universidade de S\~ao Paulo, 
CP 369, S\~ao Carlos 13560-970, S\~ao Paulo, Brazil}

\affil{
$^3$Laboratory of Information Technologies, 
Joint Institute for Nuclear Research, \\
Dubna 141980, Russia } 

\email{yukalov@theor.jinr.ru}

\begin{abstract}
By pumping energy into a trapped Bose-Einstein condensate it is possible to generate
nonlinear coherent modes representing non-ground-state condensates. A Bose-condensed 
system of trapped atoms with nonlinear coherent modes is analogous to a finite-level 
atom considered in optics which can be excited by applying external fields. The 
excitation of finite-level atoms produces a variety of optical phenomena. In the similar 
way, the generation of nonlinear coherent modes in a trapped condensate results in many 
phenomena studied in what is termed atom optics. For example, there occur such effects 
as interference patterns, interference current, Rabi oscillations, harmonic generation, 
parametric conversion, Ramsey fringes, mode locking, and a dynamic transition between 
Rabi and Josephson regimes. The possibility of creating mesoscopic entangled states of 
trapped atoms and entanglement production by atomic states in optical lattices are studied.   
\end{abstract}

\section{Introduction}

In trapped Bose-Einstein condensates, it is possible to generate nonlinear coherent 
modes by transferring atoms from the ground state to the excited states of the atomic 
cloud. The macroscopic occupation of the excited states of trapped atoms represents 
non-ground-state Bose-Einstein condensates also called nonlinear coherent modes. These
modes are nonlinear because of the interaction between atoms, and they are coherent, 
since Bose-Einstein condensates correspond to coherent states. A system of trapped
Bose-condensed atoms with nonlinear coherent modes is analogous to a finite-level atom
studied in optics \cite{Allen_1}. Moreover, since Bose-Einstein condensates correspond
to coherent states, they exhibit the properties similar to those observed in coherent
optics \cite{Mandel_2}. 

The layout of this article is as follows. In Sec. 2, we recall that the equation for the
condensate wave function represents the system coherent state. In Sec. 3, the equations 
for nonlinear coherent modes are obtained. The occurrence of the coherent states leads to 
the appearance of several interference effects discussed in Sec. 4. The generation of 
mesoscopic entangled states and entanglement production in optical lattices are also 
described in Sec. 4. Section 5 concludes.

\section{Equation for coherent field}

Let us consider a system of Bose atoms represented by the field operator $\psi(\bfr,t)$.
The Heisenberg equation of motion reads as
\be
\label{1}
 i\; \frac{\prt}{\prt t} \; \psi(\bfr,t) = H[\; \psi\; ] \; \psi(\bfr,t) \;  ,
\ee
with the nonlinear Hamiltonian
\be
\label{2}
 H[\; \psi\; ] = - \; \frac{\nabla^2}{2m} + U(\bfr,t) + 
\int \psi^\dgr(\bfr') \; \Phi(\bfr-\bfr')\; \psi(\bfr') \; d\bfr '\;   .
\ee
The external potential $U({\bf r},t)$ includes a trapping potential and its temporal 
modulation, if it exists. The interaction potential can have any form, provided it is 
integrable. Here and in what follows, the Planck constant $\hbar$ is set to one. 

The existence of a Bose-Einstein condensate implies the occurrence of a coherent state
representing the condensed fraction of the system \cite{Yukalov_3,Yukalov_4,Yukalov_5}. 
All the system can be in a coherent state if temperature is zero and atomic interactions 
are asymptotically weak. The existence of a coherent state tells us that there is a 
nontrivial eigenvalue of the field operator, called the coherent field, such that
\be
\label{3}
 \psi(\bfr,t) \; | \; \eta \; \rgl = \eta(\bfr,t) \; | \; \eta \; \rgl \;  ,
\ee
where the coherent field $\eta({\bf r},t)$ represents the condensate wave function. Then 
the averaging of equation (\ref{1}) over the coherent state $|\eta\rangle$ gives the 
equation for the coherent field 
\be
\label{4}
 i\; \frac{\prt}{\prt t} \; \eta(\bfr,t) = H[\; \eta\; ] \; \eta(\bfr,t) \;  ,
\ee
with the nonlinear Hamiltonian
\be
\label{5}
H[\; \eta\; ] = - \; \frac{\nabla^2}{2m} + U(\bfr,t) + 
\int \Phi(\bfr-\bfr')\;  |\; \eta(\bfr',t) \; |^2 \; d\bfr '\;    ,
\ee
hence the equation for the condensate wave function. The latter is normalized to the total
number of particles,
\be
\label{6}
\int  |\; \eta(\bfr,t) \; |^2 \; d\bfr  = N \;   .
\ee
  
This equation (\ref{4}) was advanced by Bogolubov \cite{Bogolubov_6} in 1949 and since then
republished numerous times (e.g. \cite{Bogolubov_7,Bogolubov_8,Bogolubov_9}). An analysis of 
possible solutions was given by Gross \cite{Gross_10,Gross_11,Gross_12,Gross_13,Gross_14}
(see also \cite{Wu_15,Pitaevskii_16}). The general mathematical structure of equation 
(\ref{4}) is that of nonlinear Schr\"{o}dinger (NLS) equation \cite{Malomed_17}.
      
In order to pass to the coherent field normalized to one, it is possible to introduce a
function $\varphi$, so that
\be
\label{7}
 \eta(\bfr,t) = \sqrt{N} \; \vp(\bfr,t) \;   ,
\ee
with the normalization 
\be
\label{8}
 \int  |\; \vp(\bfr,t) \; |^2 \; d\bfr  = 1 \;   .
\ee
Then equation (\ref{4}) reads as
\be
\label{9}
i\; \frac{\prt}{\prt t} \; \vp(\bfr,t) = \left\{ -\; \frac{\nabla^2}{2m} + U(\bfr,t) +
N \int \Phi(\bfr-\bfr') \; |\; \vp(\bfr',t) \; |^2 \; d\bfr ' \right\} \; \vp(\bfr,t) \; .
\ee

Sometimes, one considers the Schr\"{o}dinger representation, where the system wave function
satisfies the Schr\"{o}dinger equation 
\be
\label{10}
i\; \frac{\prt}{\prt t} \; \Psi_N =  H_N \Psi_N \;  ,
\ee
with the Hamiltonian
\be
\label{11}
 H_N = \sum_{i=1}^N \left[ \; -\; \frac{\nabla^2}{2m} + U(\bfr,t) \; \right] +
\frac{1}{2} \sum_{i\neq j}^N \Phi(\bfr_i-\bfr_j) \;   .
\ee
Looking for the function of the form
\be
\label{12}
\Psi_N = \prod_{i=1}^N \vp(\bfr_i,t) \;   ,
\ee
one gets the average energy
$$
\lgl \; H_N \; \rgl \equiv ( \Psi_N, \; H_N \Psi_N ) = 
N \int \left[ \; -\; \frac{\nabla^2}{2m} + U(\bfr,t) \; \right] \; 
|\; \vp(\bfr,t) \; |^2 \; d\bfr \; +
$$
\be
\label{13}
+ \;
 \frac{1}{2} \; N ( N - 1) \int |\; \vp(\bfr,t) \; |^2 \; \Phi(\bfr-\bfr') \;
|\; \vp(\bfr',t) \; |^2 \; d\bfr d\bfr ' \;   .
\ee
From the equation 
\be
\label{14}
i\; \frac{\prt}{\prt t} \; \vp(\bfr,t) = 
\frac{\dlt \lgl H_N \rgl}{\dlt \vp^*(\bfr,t) } \;
\ee
one obtains the equation of the same form as (\ref{9}), because of which one states 
that the coherent-field equation (\ref{9}) corresponds to the mean-field approximation 
(\ref{12}).

However the coincidence of the equations is rather occasional, happening only at zero
temperature and asymptotically weak interactions. For the genuine Bose-Einstein condensation,
spontaneous global gauge symmetry breaking is required \cite{Yukalov_4,Yukalov_18}. In 
the Schr\"{o}dinger approach, with a finite number of atoms, there is no this symmetry 
breaking and no phase transition, so that it is admissible to talk only on quasi-condensate. 

Moreover, the product wave function (\ref{12}) does not exhaust the mean-field approximation.
A general wave function in the Schr\"{o}dinger representation has to be constructed of 
many symmetrized orbitals \cite{Alon_19}. And in the Heisenberg representation, the general 
mean-field approximation is the Hartree-Fock-Bogolubov approximation.   

In the Heisenberg representation, the global gauge symmetry breaking is, most conveniently, 
introduced by means of the Bogolubov shift
\be
\label{15}
\psi(\bfr,t) = \eta(\bfr,t) + \psi_1(\bfr,t) \;   ,
\ee
where $\eta$ is the coherent field, or condensate wave function, and $\psi_1$ is the 
operator of uncondensed particles. The contribution of the latter diminishes with 
lowering temperature and weakening interactions. When the influence of temperature and 
interactions are negligible, we come to the coherent approximation with equations 
(\ref{4}) and (\ref{9}).

\section{Nonlinear coherent modes}

The external potential can consist of two terms, a stationary trapping potential and an
additional modulating potential,
\be
\label{16}
 U(\bfr,t) = U(\bfr) + V(\bfr,t) \;  .   
\ee
Then the Hamiltonian (\ref{5}) is also the sum of two terms,
\be
\label{17}
 H[\; \eta \; ] = \hat H[\; \vp\; ] + V(\bfr,t) \;  ,
\ee
where the first term corresponds to the stationary part
$$
\hat H[\; \vp \; ] \equiv - \; \frac{\nabla^2}{2m} + U(\bfr) + N 
\int \Phi(\bfr-\bfr') \; |\; \vp(\bfr',t) \; |^2 \; d\bfr' \;  .
$$
The eigenproblem for the stationary part,
\be
\label{18}
\hat H[\; \vp_n \; ] \; \vp_n(\bfr) = E_n \; \vp_n(\bfr) \;  ,
\ee
defines the nonlinear coherent modes $\varphi_n({\bf r})$. The solution to the 
time-dependent equation (\ref{9}) can be searched in the form of an expansion over 
the coherent modes,
\be
\label{19}
 \vp(\bfr,t) = \sum_n c_n(t) \; \vp_n(\bfr) \; e^{-iE_n t} \;  ,
\ee
with the coefficients varying in time slower than the exponential $\exp(-i E_n t)$.    

In order to excite the upper coherent modes, one can act on the system by an alternating 
external field with a modulation frequency $\omega$ in resonance with a transition 
frequency between two modes
\be
\label{20}
\om_{mn} \equiv E_m - E_n \;   ,
\ee
so that the detuning be small,
\be
\label{21}
 \left| \; \frac{\Dlt_{mn}}{\om} \; \right| \ll 1 \; , \qquad 
\Dlt_{mn} \equiv \om - \om_{mn} \;  ,
\ee
for instance transferring atoms from the ground state to a higher mode.

Thus by acting on trapped atoms, interacting through a local potential
\be
\label{22}
 \Phi(\bfr) = \Phi_0 \; \dlt(\bfr) \; , \qquad 
\Phi_0 \equiv 4\pi \; \frac{a_s}{m} \; ,
\ee
with the external potential
\be
\label{23}
V(\bfr,t) = V_1(\bfr) \cos(\om t) + V_2(\bfr) \sin(\om t) \; ,
\ee
excites the corresponding resonant mode by the action of the interaction amplitude
\be
\label{24}
\al_{mn} \equiv 
N \Phi_0 \int | \; \vp_m(\bfr) \; |^2 \left\{ 2\; | \; \vp_n(\bfr) \; |^2 - 
| \; \vp_m(\bfr) \; |^2 \right\} \; d\bfr
\ee
and the pumping-field amplitude
\be
\label{25}
\bt_{mn} \equiv 
 \int | \; \vp_m^*(\bfr) \; [ \; V_1(\bfr)  - i V_2(\bfr) \; ] \; 
\vp_n(\bfr) \; d\bfr   .
\ee
The resonant generation leads \cite{Yukalov_20,Yukalov_21} to the equations
$$
i\; \frac{d c_1}{dt} = \al_{12} |\; c_2 \; |^2 c_1 + 
\frac{1}{2} \; \bt_{12} \; c_2 \; e^{i\Dlt_{12} t} \; ,
$$
\be
\label{26}
i\; \frac{d c_2}{dt} = \al_{21} |\; c_1 \; |^2 c_2 + 
\frac{1}{2} \; \bt_{12}^* \; c_1 \; e^{- i\Dlt_{12} t} \;    .
\ee

In the similar way, by acting on atoms with two modulating potentials, with the 
frequencies $\omega_1$ and $\omega_2$, it is possible to generate two higher modes 
employing either the cascade generation, or $\Lambda$-type generation, or $V$-type 
generation, as is illustrated in Fig. 1. 

\begin{figure}[ht]
\begin{center}
\includegraphics[width=5cm]{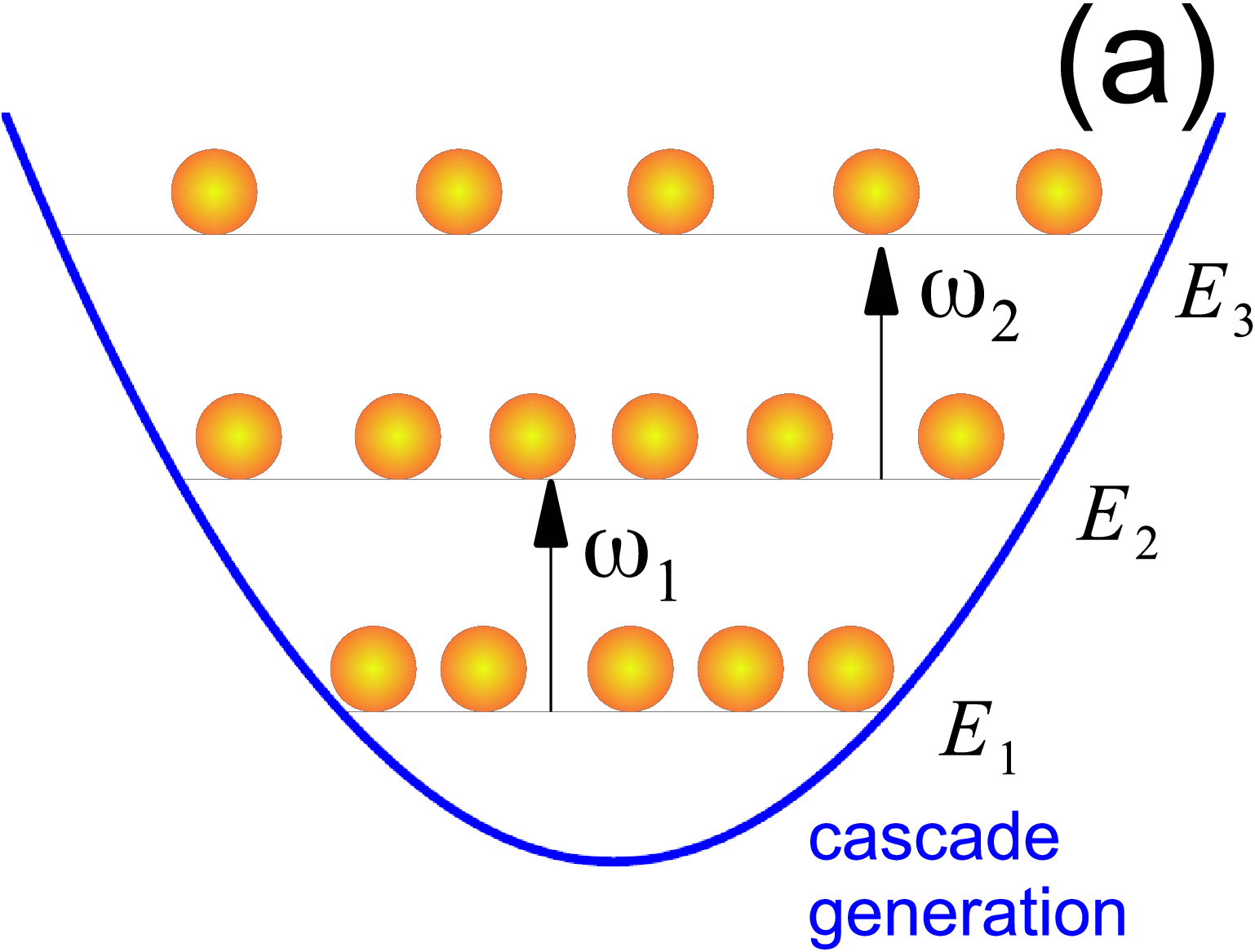} \hspace{1.5cm}
\includegraphics[width=5cm]{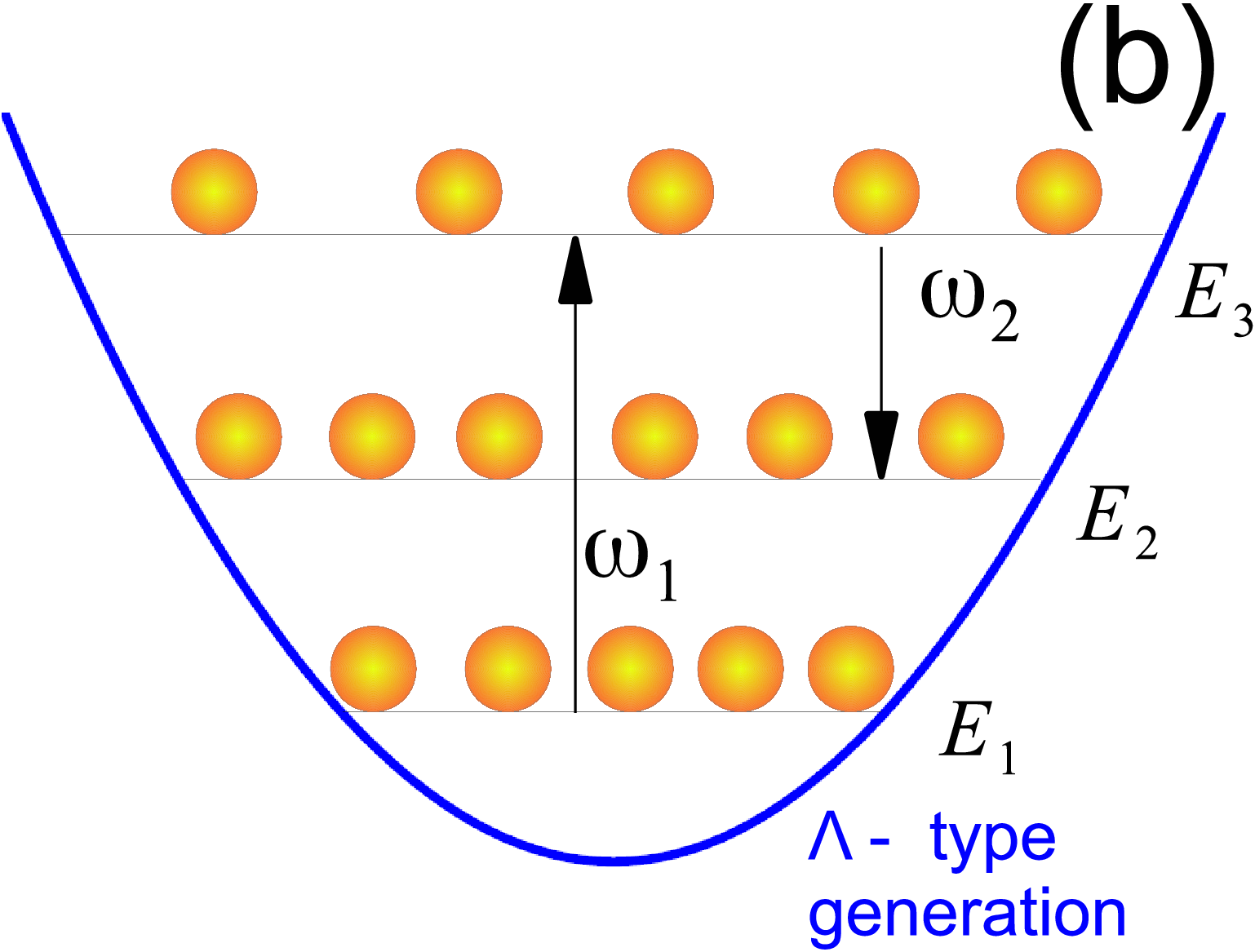}  
\end{center}
\vskip 1cm
\begin{center}
\includegraphics[width=5cm]{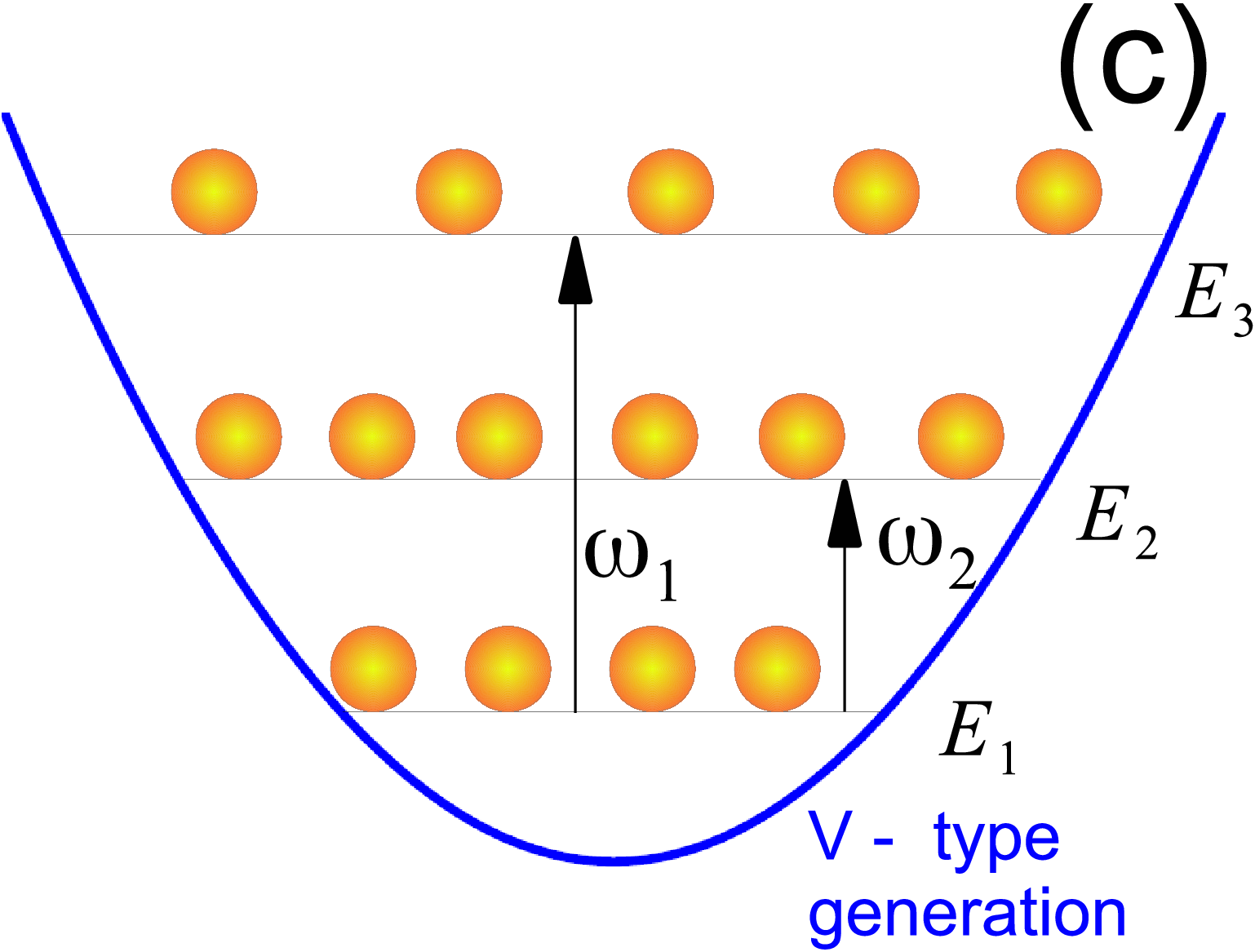} 
\end{center}
\caption{ \small
(a) Coexistence of three nonlinear coherent modes obtained
by means of cascade generation; 
(b) Three nonlinear coherent modes obtained by means of $\Lambda-$type
generation;
(c) Three nonlinear coherent modes obtained by means of $V-$type
generation.
}
\label{fig:Fig.1}
\end{figure}

\section{Coherent atom optics}

Generating nonlinear coherent modes of trapped atomic Bose-Einstein condensates allows 
for the observation of several effects some of which are analogous to optical phenomena.
The principal difference is the existence of atomic interactions making the related 
equations nonlinear. Nevertheless, in many cases it is admissible to find approximate 
solutions resorting to the scale separation approach \cite{Yukalov_23,Yukalov_24}
stemming from the Krylov-Bogolubov averaging techniques \cite{Krylov_25,Bogolubov_26}.
These methods also allow for considering random noise as a fast variable 
\cite{Horsthemke_27}, hence the influence of noise on the dynamics of nonlinear coherent 
modes can also be taken into account \cite{Yukalov_28}. Below we list several effects
that we have considered concentrating on the results, while the details can be found
in the cited papers and summarized in the review article \cite{Yukalov_22}.

\subsection{Interference patterns}

First of all, since the nonlinear modes are coherent, they form interference patterns 
that can be observed by measuring the atom density in a trap \cite{Yukalov_21}. The 
interference term is the difference
\be
\label{27}
\rho_{int}(\bfr,t) = \rho(\bfr,t) - \sum_n \rho_n(\bfr,t)
\ee
between the total atomic density and the sum of the mode densities
\be
\label{28}
 \rho_n(\bfr,t) = N \; | \; c_n(t) \; \vp_n(\bfr) \; |^2 \;  ,
\ee
which gives
\be
\label{29}
\rho_{int}(\bfr,t) = N \sum_{m\neq n} c_m^*(t) \; c_n(t) \; 
\vp_m^*(\bfr) \; \vp_n(\bfr) \; e^{i\om_{mn} t} \;   .
\ee

\subsection{Interference current}

Similarly \cite{Yukalov_21}, the difference 
\be
\label{30}
{\bf j}_{int}(\bfr,t) = {\bf j}(\bfr,t) - \sum_n {\bf j}_n(\bfr,t) 
\ee
between the total current in the system and the sum of the mode currents
\be
\label{31}
{\bf j}_n(\bfr,t) = \frac{N}{m} \; {\rm Im} \; |\; c_n(t)\;|^2 \; \vp_n(\bfr) \;
\nabla \vp_n(\bfr)
\ee
defines the interference current
\be
\label{32}
{\bf j}_{int}(\bfr,t) =  \frac{N}{m} \; {\rm Im} \sum_{m\neq n}  
c_m^*(t)\; c_n(t) \; \vp_m^*(\bfr) \;
\nabla \vp_n(\bfr) \; e^{i\om_{mn} t} \;  .
\ee

\subsection{Rabi oscillations}

The fractional mode populations, starting with the initial conditions for the coefficient 
functions
\be
\label{33}
c_1(0) = 1 \; , \qquad c_2(0) = 0 \;   ,
\ee
oscillate \cite{Yukalov_20} following the law
\be
\label{34}
n_1 = 1 - \; \frac{|\;\bt_{12}\;|^2}{\Om^2} \; \sin^2\left( \frac{\Om t}{2} \right) \; ,
\qquad
n_2 =  \frac{|\;\bt_{12}\;|^2}{\Om^2} \; \sin^2\left( \frac{\Om t}{2} \right) \;
\ee
with the effective Rabi frequency
\be
\label{35}
 \Om^2 = |\; \Dlt_{eff} \; |^2 + |\; \bt_{12} \; |^2 \;  ,
\ee
where the effective detuning is
\be
\label{36}
 \Dlt_{eff} \equiv \Dlt_{21} + \al_{12} n_2 - \al_{21} n_1 \qquad
(\Dlt_{21} \equiv \om - \om_{21} ) \;   .
\ee
These oscillations remind the standard Rabi oscillations in optics \cite{Rabi_29},
however, these are not simple sign oscillations, as far as the effective Rabi frequency
(\ref{35}) includes time dependent fractional mode populations $n_i=n_i(t)$.

\subsection{Mode locking}

In the dynamics with two modes, there exists a regime \cite{Yukalov_20,Yukalov_21}
where the fractional mode populations are locked in a limited range of their admissible 
values, so that
\be
\label{37}
 \frac{1}{2} < n_1 \leq 1 \; , \qquad 0 \leq n_2 < \frac{1}{2} \qquad
\left(   |\; \bt_{12} \; | + \Dlt_{21} < \frac{1}{2} \; \al_c \right) \; ,
\ee
where
\be
\label{38}
 \al_c \equiv \frac{1}{2} \; ( 3\al_{21} - \al_{12} ) \;  .
\ee

Increasing the pumping amplitude or detuning shifts the dynamics of the mode populations
into the whole admissible range between $0$ and $1$, 
\be
\label{39}
0 \leq n_i \leq 1 \qquad 
\left(   |\; \bt_{12} \; | + \Dlt_{21} > \frac{1}{2} \; \al_c \right) \;  ,
\ee
so that the modes become unlocked.

\subsection{Rabi-Josephson transition}

The mode locked regime is termed the Rabi regime, while the mode unlocked regime is called
Josephson regime. On the parameter line
\be
\label{40}
   |\; \bt_{12} \; | + \Dlt_{21} = \frac{1}{2} \; \al_c  \;  ,
\ee
there occurs the dynamic transition between the Rabi and Josephson regimes. Averaging the 
Hamiltonian over time results in a stationary system that exhibits on the critical line
(\ref{40}) a phase transition of second order \cite{Yukalov_30}.

\subsection{Ramsey fringes}

Similarly to Ramsey fringes in optics \cite{Ramsey_31}, Ramsey fringes in a trap with 
nonlinear coherent modes \cite{Yukalov_22,Ramos_32,Ramos_33} characterize the excited 
mode fractional population $n_2 = |c_2|^2$ after the action of two consecutive $\pi/2$  
pulses of the pumping field, of the temporal length $\tau$, separated by a long time 
interval $T \gg \tau$. Keeping in mind the initial conditions (\ref{33}), we have
\be
\label{41}
n_2(2\tau + T) = 
\frac{|\;\bt_{12}\;|^2}{\Om^2} \; \left[ \; \cos\left( \frac{\Om \tau}{2} \right)  + 
\frac{\Dlt_{eff}}{\Om} \; \sin\left( \frac{\Dlt_{eff} T}{2} \right) \; \right ] \; ,
\ee
where $\Omega \tau = \pi/2$. This looks like the usual Ramsey expression in optics. However,
the effective detuning, as well as the effective Rabi frequency depend on time through the
mode populations $n_i = n_i(t)$.

\subsection{Chaotic dynamics}

When there coexist two coherent modes, the oscillations can be rather complicated due to
the nonlinearities caused by atom interactions, but anyway they do not display chaotic 
behavior. However when three modes coexist \cite{Yukalov_34,Yukalov_35}, the populations 
can either oscillate or they can exhibit chaotic behavior at sufficiently large pumping 
amplitude. Setting, for simplicity $\alpha_{mn}=\alpha$ and $\beta_{mn} = \beta$, we find 
that chaotic behavior appears when
\be
\label{42}
\left| \; \frac{\bt}{\al} \; \right| \geq 0.639448 \;   .
\ee

\subsection{Higher-order resonances}

In addition to the resonance conditions where the frequencies of the modulating fields
are close to interlevel transition frequencies, there exist higher-order resonances 
\cite{Yukalov_34,Yukalov_35}, such as harmonic generation, when the frequency of the 
modulating field satisfies the relation
\be
\label{43}
n\om = \om_{21} \qquad ( n = 1,2, \ldots ) \;   .
\ee
In the presence of several frequencies $\omega_i$ of alternating fields there happens
parametric conversion, when
\be
\label{44}
\sum_j ( \pm \om_j ) = \om_{21} \;   .
\ee

\subsection{Atomic squeezing}

Since atomic traps are finite systems, strictly speaking, there develops not a genuine 
condensate but a quasi-condensate that can be characterized by the mode creation and 
destruction operators. It is possible to introduce pseudospin operators $S^\alpha$, 
with $\alpha = x,y,z$. In this representation, the operator $S^z$ corresponds to the mode 
population difference, while the ladder operators $S^{\pm}$ describe intermode transitions.
The squeezing of the operator $S^z$ with respect to $S^{\pm}$ is quantified by the squeezing
factor 
\be
\label{45}
 Q(S^z , \; S^\pm ) = \frac{2{\rm var}(S^z)}{|\; \lgl\; S^\pm\; \rgl \; | } \;  ,
\ee
where the operator variance is
$$
{\rm var}(\hat A)  \equiv \lgl \; \hat A^+ \;\hat A \; \rgl -
|\; \lgl \; \hat A \; \rgl \; |^2 \; .
$$
In the case of two modes, we find \cite{Yukalov_21,Yukalov_22}
\be
\label{46}
Q(S^z , \; S^\pm ) = \sqrt{1 - s^2 } \;   ,
\ee
with the population difference
\be
\label{47}
 s = \frac{2}{N} \;  \lgl \; S^z \; \rgl = n_2 - n_1 \; 	 .
\ee
As far as $s \leq 1$, the operator $S^z$ is almost always squeezed with respect to $S^{\pm}$.
This implies that the population difference can be measured more accurately than the phase
difference or atomic current.

\subsection{Mesoscopic entanglement}

By generating nonlinear coherent modes in optical lattices, with atomic clouds in each 
lattice site, it is possible to create mesoscopic entangled states \cite{Yukalov_36}.
Suppose, Bose-condensed atoms are loaded into a deep optical lattice, where at each 
lattice site there are many atoms. The lattice sites are enumerated with the index
$j=1,2,\ldots,N_L$. By shaking the lattice uniformly, with the same resonance condition for
all sites, it is possible to generate the same mode in each lattice site, as is shown in 
Fig. 2. Let the modes be enumerated by $m=1,2,\ldots,M$. At each lattice site, there are 
$N$ Bose-condensed atoms, with $N \gg 1$. Experiments \cite{Hadzibabic_37,Cennini_38} 
prove that each lattice site can really house quite a number of Bose-condensed atoms, 
of order $10^4$. 

\begin{figure}[ht]
\begin{center}
\includegraphics[width=8cm]{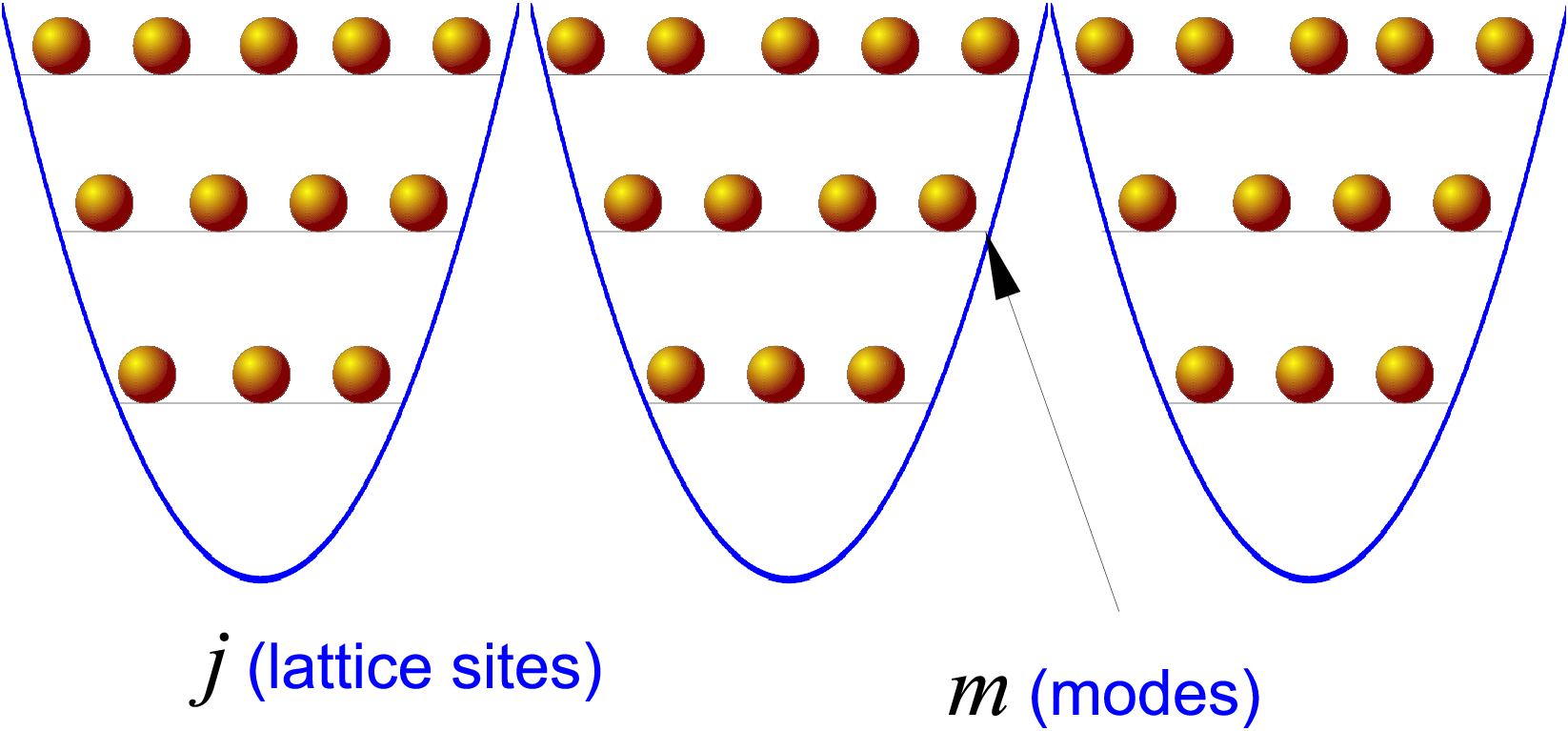} 
\end{center}
\vskip 1cm
\caption{ \small
Mesoscopic entangled state of atomic clouds in different lattice sites of
an optical lattice. Multimode multicat state.
}
\label{fig:Fig.2}
\end{figure}

A coherent mode $m$ at a lattice site $j$, which is characterized by a coherent field
$\eta_{jm}({\bf r})$, in the Fock space is represented \cite{Yukalov_39} by a column   
\be
\label{48}
|\; \eta_{jm} \; \rgl = \left[ \;
\frac{e^{-N/2}}{\sqrt{n!} } \; \prod_{i=1}^n \eta_{jm}(\bfr_i) \; \right] \;   ,
\ee
where the rows are enumerated by $n=1,2,\ldots$. The closed linear envelope of these modes
forms the Hilbert space 
\be
\label{49}
 \cH_j = \overline \cL_m \{ \; |\; \eta_{jm}\; \rgl \; \} \;  .
\ee
The state of the overall system of $N_L$ lattice wells, with an excited mode $m$ is
\be
\label{50}
|\; \eta_m \; \rgl = \bigotimes_{j=1}^{N_L} |\; \eta_{jm}\; \rgl \;  .
\ee
Then the whole system Hilbert space is
\be 
\label{51}
\cH = \overline \cL_m \{ \; |\; \eta_{m}\; \rgl \; \} =
\bigotimes_{j=1}^{N_L} \cH_j \;    .
\ee
The states of space (\ref{51}) are
\be
\label{52}
|\; \eta \; \rgl = \sum_{m=1}^M c_m(t) \; | \; \eta_m \; \rgl \; ,
\ee
so that the statistical operator is
\be
\label{53}
 \hat\rho(t) = |\; \eta \; \rgl \lgl \; \eta \; | \;   .
\ee
This operator has the form
\be
\label{54}
 \hat\rho(t) = \sum_{mn}^M c_m(t) \; c_n^*(t) \; \bigotimes_{j=1}^{N_L}
|\; \eta_{jm} \; \rgl \lgl \; \eta_{jn} \; | \;   ,   
\ee
which shows that it is an entangled multimode multicat state, provided that there are 
at least two modes, hence $M \geq 2$. The operator (\ref{54}) describes a state where 
the mesoscopic atomic clouds in different lattice wells are entangled between each other.

\subsection{Entanglement production}

The statistical state (\ref{54}) acting even on the not entangled vector states can 
transform them into entangled states. This is important in calculating observable 
quantities defined as the averages of Hermitian operators. The entangling power of a
trace-class operator $\hat{A}$, such that
\be
\label{55}
0 < |\; {\rm Tr} \; \hat A \; | < \infty \;   ,
\ee
acting on a Hilbert space of form (\ref{51}), is defined as follows 
\cite{Yukalov_40,Yukalov_41,Yukalov_42}.

The action of a considered operator $\hat{A}$ is compared with that of its nonentangling 
counterpart
\be
\label{56}
\hat A^\otimes \equiv 
\frac{\bigotimes_j \hat A_j}{({\rm Tr}_\cH \hat A)^{N_L-1} } \;   ,
\ee
in which 
\be
\label{57}
\hat A_j \equiv {\rm Tr}_{\cH/\cH_i} \hat A
\ee
is a partially traced operator and the normalization condition is imposed,
\be
\label{58}
{\rm Tr}_\cH \hat A^\otimes =  {\rm Tr}_{\cH} \hat A \;   .
\ee
The tensor-product operators do not produce entanglement 
\cite{Marcus_46,Westwick_47,Beasley_48,Alfsen_49,Johnston_50,Friendland_51}. 

The measure of entanglement production of the operator $\hat{A}$ is defined 
\cite{Yukalov_40,Yukalov_41,Yukalov_42} as
\be
\label{59}
 \ep(\hat A )  \equiv 
\log \; \frac{||\; \hat A \;||}{||\; \hat A^\otimes \;||} \; .
\ee
This measure is semi-positive, zero for nonentangling operators, continuous, additive, 
and invariant under local unitary operations.

Then the measure of entanglement production of the statistical operator (\ref{54}),
\be
\label{60}
 \ep(\hat\rho(t) ) = 
\log \; \frac{||\; \hat\rho(t) \;||}{||\; \hat\rho^\otimes(t) \;||}  \;  ,
\ee
using the standard operator norm, yields \cite{Yukalov_43}
\be
\label{61}
 \ep(\hat\rho(t) ) = ( 1 - N_L) \; \log\; \sup_m \; n_m(t) \;   ,
\ee
where $n_m(t) = |c_m(t)|^2$ is the fractional population of a mode $m$. Other examples 
of entanglement production are given in \cite{Yukalov_44,Yukalov_45}

\section{Conclusion}

In trapped Bose-Einstein condensates, it is possible to generate nonlinear coherent modes
representing non-ground-state Bose condensates. They are called nonlinear since atoms 
interact with each other, and the modes are coherent, as far as Bose-Einstein condensates 
are coherent systems. Generally, the nonlinear coherent modes can be generated either by
employing weak modulation by an external field that is in resonance with a transition 
frequency, or by a strong pumping field that does not need to be resonant \cite{Yukalov_22}.
In the case of the resonant generation, the coexistence of pure modes can be supported only 
for a finite time, after which the effect of power broadening comes into play and other 
energy levels become involved, so that good resonance cannot be anymore realized. This 
resonance time is estimated to be of the order of $10$ s, which is about the lifetime of
atoms in a trap.        

Trapped Bose condensates with nonlinear coherent modes are similar to finite-level atoms, 
and they exhibit many phenomena analogous to those occurring in coherent optics because 
of which the branch of physics studying these phenomena is named coherent atom optics. 
In the present communication, we have given a survey of the main properties of the 
Bose-Einstein condensates with nonlinear coherent modes, paying the major attention to 
the problems and effects investigated by the authors. In particular, we have considered 
interference patterns, interference current, Rabi oscillations, mode locking, Rabi-Josephson 
transitions, Ramsey fringes, chaotic dynamics, higher-order resonances, resulting in 
harmonic generation and parametric conversion, atomic squeezing, mesoscopic entanglement, 
and entanglement production. More information on these and other properties of trapped 
Bose-Einstein condensates with nonlinear coherent modes can be found in the recent 
review \cite{Yukalov_22}.

\section*{Author Contributions}

All the authors, V.I. Yukalov, E.P. Yukalova, and V.S. Bagnato equally contributed to this 
work.

\vskip 5mm
{\parindent=0pt
This research did not receive any specific grant from funding agencies in the public, 
commercial, or not-for-profit sectors. }

\end{document}